\documentclass[fp,preprint]{jpsj3}
\usepackage{txfonts}
\usepackage[usenames, dvipsnames]{color}
\usepackage{epstopdf}
\newcommand{\beq}{\begin{eqnarray}}
\newcommand{\eeq}{\end{eqnarray}}

\title{Implementation of Generalized Bloch Theorem Using Linear Combination of Pseudo-Atomic Orbitals}

\author{Teguh Budi Prayitno$^{1,2}$\thanks{teguh@cphys.s.kanazawa-u.ac.jp, teguh-budi@unj.ac.id} and Fumiyuki Ishii$^3$\thanks{ishii@cphys.s.kanazawa-u.ac.jp}}
\inst{$^1$Graduate School of Natural Science and Technology, Kanazawa University, Kanazawa 920-1192, Japan \\
$^2$Department of Physics, Faculty of Mathematics and Natural Science, Universitas Negeri Jakarta, Kampus A Jl. Rawamangun Muka, Jakarta Timur, Indonesia\\
$^3$Nanomaterials Research Institute, Kanazawa University, Kanazawa 920-1192, Japan}

\abst{We have implemented the generalized Bloch theorem based on first-principles calculations using a linear combination of pseudo-atomic orbitals (LCPAO) as the basis sets. In order to test our implementation in a code, we examined three systems that have been reported in experiments or other calculations, namely, the carrier-induced spin-spiral ground state in the one-dimensional model system, the spin stiffness of bcc-Fe, and the spin stiffness in a zigzag graphene nanoribbon. We confirmed that our implementation gives good agreement with experiments. Based on these results, we believe that our implementation of the generalized Bloch theorem using an LCPAO is useful for predicting the properties of complex magnetic materials. We also predicted a large reduction (enhancement) of spin stiffness for the electron (hole) doping of zigzag-edge graphene nanoribbon ferromagnetic states.
}

\begin{document}
\maketitle

\section{Introduction}
\indent Noncollinear magnetic structures, which are extensions of the collinear magnetic structures, are of interest for the in-depth study of the hidden phenomena in condensed matter physics. The structure of noncollinear magnetism has been discovered experimentally for crystalline and amorphous materials \cite{0Coey}. Theoretically, for the whole crystal, a noncollinear magnetic structure has no natural spin quantization axis, contrary to the collinear magnetic structure. Therefore, the magnetic moment of each atom can have a different direction in the crystal. A typical noncollinear magnetic structure is the spin spiral, where each magnetic moment is rotated around a specific spin rotation axis from atom to atom on a periodic lattice by a constant angle following a certain direction of the crystal.    
    
\indent Due to the translation symmetry breaking in a Bravais lattice, to consider a spiral structure by first-principles calculation, especially in the case of long-wavelength spin waves, we have to make a large unit cell. By neglecting spin-orbit coupling (SOC), this problem can be solved using the generalized Bloch theorem (GBT) by considering the minimal number of atoms in the primitive unit cell \cite{1Sandratskii}. In other words, without SOC, we are allowed to calculate a spiral system efficiently without using a large unit cell by implementing the GBT. In addition to its advantage, by using first-principles calculation, the implementation of the GBT can also be used to deal with the magnon dispersion relation and its related quantities, such as computing the exchange coupling constant and Curie temperature \cite{2Rosengaard,3Halilov,4Padja, 7.1Jakobson, 7.2Jakobson}, the Dzyaloshinskii$\textendash$Moriya interaction including SOC perturbatively \cite{5Heide}, and so forth.  

\indent We have implemented the GBT in the framework of a linear combination of pseudo-atomic orbitals (LCPAO) as basis sets, as implemented in the OPENMX code \cite{6Openmx}. There are some advantages of using an LCPAO as basis sets, especially in the implementation of density functional theory (DFT). Due to its localized property, an LCPAO can be employed to treat large systems efficiently using the so-called O($N$) method as implemented in DFT calculation \cite{Yang, Ozaki-Terakura}. In addition, good accuracy in describing electronic properties can be obtained for large systems by using only a minimal basis set \cite{Portal, Lewis}. 

The combination of the GBT with an LCPAO as basis sets has an advantage over the GBT with plane-wave basis sets. The advantage of using an LCPAO for the GBT is related to the computational time, especially for calculating spin stiffness. Evaluating spin stiffness needs a large amount of $k$ point sampling and many spiral vectors $\textit{\textbf{q}}$ around $\textit{\textbf{q}}=0$. The problem of using plane-wave basis sets arises when treating the vacuum region for a one or two-dimensional system. Treating the vacuum region requires more plane waves than an LCPAO, so using an LCPAO is more efficient than using plane waves. This means that more complex systems including a vacuum region can be handled by an LCPAO more efficiently than by using plane waves.  

Regarding the O($N$) method, we clarify that this method and the GBT can be treated as complementary methods. Using the OPENMX code, the O($N$) method has been successfully implemented to treat large systems \cite{OzakiO, Ohwaki}. However, the computational cost is still very expensive, especially for investigating long-wavelength spin waves by supercell calculation. The GBT minimizes the computational cost by using the minimal number of atoms for long-wavelength spin waves. However, the GBT can only handle homogeneous spin spirals, while the O($N$) method can handle inhomogeneous ones. Therefore, the choice of using either the O($N$) method or the GBT depends on the order of the magnetic moment.    

It is known that the spin stiffness is very important for predicting the Curie temperature, which can be computed by using the mean field approximation (MFA) or random phase approximation (RPA) \cite{4Padja}. To calculate the spin stiffness using the frozen magnon method with a small deviation from the ground state, constraint DFT for the spin orientation is needed \cite{16Gebauer}. The OPENMX code provides a constraint method by introducing a penalty functional in such a way that the spin orientation can be fixed. This penalty functional is generated by the difference between the initial and final density matrices to control the spin orientation self-consistently \cite{14Gebauer, 15Kurz}. The applications of the constraint scheme in the OPENMX code can be found in Refs. \citen{33nanorib4, sawada2, mizuta}.    

A similar implementation of the GBT combined with an LCPAO has also been reported previously \cite{SIESTA}. While the authors in Ref. \citen{SIESTA} successfully computed the spin-spiral ground state of fcc-Fe \cite{24Garcia}, they did not combine the GBT with the spin constraint method. Therefore, to our knowledge, there have been no attempts to compute spin stiffness by the GBT and an LCPAO. We claim that our implementation can also give accurate results compared with other implementations \cite{2Rosengaard, 16Gebauer}. In addition, if we combine the GBT and an LCPAO with other methods, such as LDA+U \cite{6Openmx, HanL1, HanL2, HanL3}, we can apply our implementation to a wide variety of complex materials. 

In this paper, to support our implementation, we have investigated the carrier-induced spin-spiral ground state and spin stiffness. As a test case, we investigated the appearance of the carrier-induced spin-spiral ground state of a one-dimensional hydrogen chain by changing the hole doping. We observed the magnetic phase transition when the hole doping was performed. Next, we evaluated the spin stiffness of bcc-Fe and the convergence in terms of the basis set. We obtained reliable results compared with both experimental results and other calculations. Finally, we investigated the enhancement or reduction of spin stiffness with respect to the amount of hole-electron doping for a zigzag graphene nanoribbon. For the nondoping case, we obtained a high value of spin stiffness, as predicted for the $p$ electron system. Based on the calculated results, we believe that our implementation using an LCPAO can treat larger and more complex magnetic materials, especially when calculating spin stiffness.           

\section{Method}

\indent This section is devoted to describing the details of including the GBT into the noncollinear spin DFT provided in the OPENMX code and to exploring the derivation of the magnon dispersion relation.       
\subsection{Formulation of the GBT for the LCPAO basis sets}
\indent  The main principle of a spin spiral is to rotate the magnetic moment of an atom from one unit cell to other unit cells following the direction of a certain spiral vector $\textit{\textbf{q}}$. Consider the rotation of the $i$th atom, which is assigned to site $\textit{\textbf{R}}_{i}$, by a constant angle $\varphi_{i}=\textit{\textbf{q}}\cdot \textit{\textbf{R}}_{i}$. Then its magnetic moment is given by
\beq
	\textit{\textbf{M}}_{i}(\textit{\textbf{r}}+\textit{\textbf{R}}_{i})=M_{i}(\textit{\textbf{r}}) \left(
\begin{array}{cc}
\cos\left(\varphi_{0}+\textit{\textbf{q}}\cdot \textit{\textbf{R}}_{i}\right)\sin\theta_{i}\\
\sin\left(\varphi_{0}+\textit{\textbf{q}}\cdot \textit{\textbf{R}}_{i}\right)\sin\theta_{i}\\
\cos\theta_{i}\end{array} 
\right). \label{moment}  
\eeq
From this equation, it is clear that one has to set the nonzero cone angle $\theta$ of the initial magnetic moment to make the GBT work. Note that $\theta$ and $\varphi$ are the polar and azimuthal angles in spherical coordinates, respectively, if we define the $z$ axis as the spin rotation axis. We remind readers that as long as the cone angle is the same, due to the decoupling between the real space and spin space, the spin spirals become equivalent even though the spin rotation axis can be different.  
\begin{figure}[h!]
\vspace{4mm}
\centering
\includegraphics[scale=1.0, width =!, height =!]{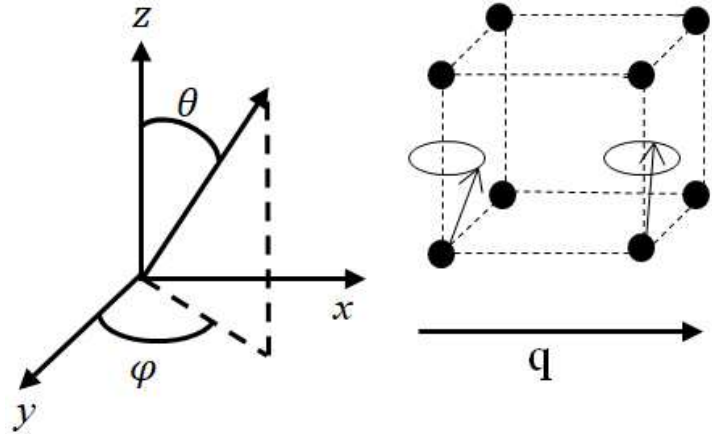}
\vspace{2mm}
\caption{\label{spiral_contoh} Illustration of a conical spiral structure with spiral vector \textit{\textbf{q}} along the $x$ axis in a cubic lattice.} 
\vspace{-4mm}
\end{figure}

The OPENMX is an open-source package used to perform material simulations adopting DFT, using norm-conserving pseudopotentials \cite{13.1Troullier}. It employs pseudo-atomic orbitals (PAO) as basis sets generated by the confinement scheme method \cite{13Ozaki, 13.2Ozaki}. To apply the GBT to the OPENMX code, we initially extend the translation operator to a generalized translation operator $\mathcal{T}_{i}$ that translates an atom and rotates its magnetic moment simultaneously from one unit cell to other unit cells following the direction of a certain spiral vector $\textit{\textbf{q}}$ as shown in Fig. \ref{spiral_contoh}. It is clear that we only introduce a new rotation operator $U(\varphi)$ that rotates a spinor in the spin space by a constant angle around the spin rotation axis. Note that if we have a Bloch wavefunction $\psi_{\textit{\textbf{k}}}(\textit{\textbf{r}})$, then the generalized translation operator applied to the Bloch wavefunction should still obey the Bloch condition
  \beq
	\mathcal{T}_{i}\psi_{\textit{\textbf{k}}}(\textit{\textbf{r}})=U(-\varphi)\psi_{\textit{\textbf{k}}}(\textit{\textbf{r}}+\textit{\textbf{R}}_{i})=e^{i\textit{\textbf{k}}\cdot\textit{\textbf{R}}_{i}}\psi_{\textit{\textbf{k}}}(\textit{\textbf{r}}),\label{bloch}
	\eeq
where the spin rotation matrix now is defined as
   \beq
U(\varphi)=\left(
 \begin{array}{cc}
e^{-i\varphi/2}& 0\\
0 & e^{i\varphi/2}\end{array} 
\right). \label{rotmat}
\eeq
In order to preserve Eq. (\ref{bloch}), it can be easily proven that the two-component spinor Bloch wavefunction should be written as
  \beq
	\psi_{\textit{\textbf{k}}}(\textit{\textbf{r}})=e^{i\textit{\textbf{k}}\cdot\textit{\textbf{r}}}\left(
\begin{array}{cc}
e^{-i\textit{\textbf{q}}\cdot\textit{\textbf{r}}/2} \alpha_{\textit{\textbf{k}}}^{\uparrow}(\textit{\textbf{r}})\\
e^{i\textit{\textbf{q}}\cdot\textit{\textbf{r}}/2} \alpha_{\textit{\textbf{k}}}^{\downarrow}(\textit{\textbf{r}})\end{array} 
\right),\label{wavefunc}
\eeq
where $\alpha_{\textit{\textbf{k}}}^{\uparrow}(\textit{\textbf{r}})$ and $\alpha_{\textit{\textbf{k}}}^{\downarrow}(\textit{\textbf{r}})$ have a translational periodicity obeying $\alpha_{\textit{\textbf{k}}}^{\sigma}(\textit{\textbf{r}})=\alpha_{\textit{\textbf{k}}}^{\sigma}(\textit{\textbf{r}}+\textit{\textbf{R}}_{i})$, in which $\sigma$ can be up or down components. In the OPENMX code, by using the definition in Eq. (\ref{wavefunc}), the Bloch wavefunction is expanded as an LCPAO $\phi_{i\alpha}$ fixed on site $\tau_{i}$ as
\beq
\psi_{\nu\textit{\textbf{k}}}\left(\textit{\textbf{r}}\right)&=&\frac{1}{\sqrt{N}}\sum_{n}^{N}\sum_{i\alpha}\left[e^{i\left(\textit{\textbf{k}}-\frac{\textit{\textbf{q}}}{2}\right)\cdot\textit{\textbf{R}}_{n}}C_{\nu\textit{\textbf{k}},i\alpha}^{\uparrow}
\left(
\begin{array}{cc}
1\\
0\end{array}
\right)+e^{i\left(\textit{\textbf{k}}+\frac{\textit{\textbf{q}}}{2}\right)\cdot\textit{\textbf{R}}_{n}}C_{\nu\textit{\textbf{k}},i\alpha}^{\downarrow}\left(
\begin{array}{cc}
0\\
1\end{array}
\right)\right]\nonumber\\
& &\times\phi_{i\alpha}\left(\mathrm{\textit{\textbf{r}}-\tau_{i}-\textit{\textbf{R}}_{n}}\right).\label{lcpao}
\eeq 
By using the density operator
\beq
\hat{n}=\sum_{\nu}^{\mathrm{occ}}f_{\nu}\left|\psi_{\nu\textit{\textbf{k}}}\right\rangle\left\langle \psi_{\nu\textit{\textbf{k}}}\right|, \label{chargeden}
\eeq
the noncollinear electron density in real space can be composed in the matrix form
  \beq
	n_{\sigma\sigma'}\left(\mathrm{\textit{\textbf{r}}}\right)&=&\left\langle \textit{\textbf{r}}\sigma\right|\hat{n}\left|\textit{\textbf{r}}\sigma'\right\rangle\nonumber\\
	&=&\left(
 \begin{array}{cc}
n_{\uparrow\uparrow}\left(\mathrm{\textit{\textbf{r}}}\right)& n_{\uparrow\downarrow}\left(\mathrm{\textit{\textbf{r}}}\right)\\
n_{\downarrow\uparrow}\left(\mathrm{\textit{\textbf{r}}}\right)& n_{\downarrow\downarrow}\left(\mathrm{\textit{\textbf{r}}}\right)\end{array} 
\right), \label{nmatrix}
	\eeq
	whose components are given as
	\beq
n_{\uparrow\uparrow}\left(\mathrm{\textit{\textbf{r}}}\right)&=&\sum_{\nu}^{\mathrm{occ}}f_{\nu}\sum_{n}^{N}e^{i\left(\textit{\textbf{k}}-\frac{\textit{\textbf{q}}}{2}\right)\cdot\textit{\textbf{R}}_{n}}\sum_{i\alpha,j\beta}C_{\nu\textit{\textbf{k}},i\alpha}^{\uparrow\ast}C_{\nu\textit{\textbf{k}},j\beta}^{\uparrow}\nonumber\\
& &\times\phi_{i\alpha}\left(\mathrm{\textit{\textbf{r}}-\tau_{i}}\right)\phi_{j\beta}\left(\mathrm{\textit{\textbf{r}}-\tau_{j}-\textit{\textbf{R}}_{n}}\right),\\
n_{\uparrow\downarrow}\left(\mathrm{\textit{\textbf{r}}}\right)&=&\sum_{\nu}^{\mathrm{occ}}f_{\nu}\sum_{n}^{N}e^{i\left(\textit{\textbf{k}}-\frac{\textit{\textbf{q}}}{2}\right)\cdot\textit{\textbf{R}}_{n}}\sum_{i\alpha,j\beta}C_{\nu\textit{\textbf{k}},i\alpha}^{\downarrow\ast}C_{\nu\textit{\textbf{k}},j\beta}^{\uparrow}\nonumber\\
& &\times\phi_{i\alpha}\left(\mathrm{\textit{\textbf{r}}-\tau_{i}}\right)\phi_{j\beta}\left(\mathrm{\textit{\textbf{r}}-\tau_{j}-\textit{\textbf{R}}_{n}}\right),\\
n_{\downarrow\uparrow}\left(\mathrm{\textit{\textbf{r}}}\right)&=&\sum_{\nu}^{\mathrm{occ}}f_{\nu}\sum_{n}^{N}e^{i\left(\textit{\textbf{k}}+\frac{\textit{\textbf{q}}}{2}\right)\cdot\textit{\textbf{R}}_{n}}\sum_{i\alpha,j\beta}C_{\nu\textit{\textbf{k}},i\alpha}^{\uparrow\ast}C_{\nu\textit{\textbf{k}},j\beta}^{\downarrow}\nonumber\\
& &\times\phi_{i\alpha}\left(\mathrm{\textit{\textbf{r}}-\tau_{i}}\right)\phi_{j\beta}\left(\mathrm{\textit{\textbf{r}}-\tau_{j}-\textit{\textbf{R}}_{n}}\right)\\
n_{\downarrow\downarrow}\left(\mathrm{\textit{\textbf{r}}}\right)&=&\sum_{\nu}^{\mathrm{occ}}f_{\nu}\sum_{n}^{N}e^{i\left(\textit{\textbf{k}}+\frac{\textit{\textbf{q}}}{2}\right)\cdot\textit{\textbf{R}}_{n}}\sum_{i\alpha,j\beta}C_{\nu\textit{\textbf{k}},i\alpha}^{\downarrow\ast}C_{\nu\textit{\textbf{k}},j\beta}^{\downarrow}
\nonumber\\
& &\times\phi_{i\alpha}\left(\mathrm{\textit{\textbf{r}}-\tau_{i}}\right)\phi_{j\beta}\left(\mathrm{\textit{\textbf{r}}-\tau_{j}-\textit{\textbf{R}}_{n}}\right).\label{compmatrix}
\eeq
In Eqs. (\ref{lcpao})-(\ref{compmatrix}), Roman and Greek indexes in the subscripts refer to the site index and orbital index, respectively. Based on the above analytical expressions, we developed a code in the OPENMX that implements the GBT.     

\subsection{General procedure for obtaining magnon dispersion relation}
 \indent In order to obtain the spin stiffness of bcc-Fe and a zigzag graphene nanoribbon, a special treatment should be considered. However, for the carrier-induced spin-spiral ground state of a hydrogen chain, we can perform a direct self-consistent calculation for a given spiral vector $\textit{\textbf{q}}$ by using the GBT. To obtain the magnon dispersion relation that will be used to calculate the spin stiffness, we use the frozen magnon method, which has been implemented widely to describe the long-wavelength excitations in itinerant ferromagnets where the magnetic moment deviates from the ground state with a small fixed cone angle $\theta$, see Fig. \ref{spiral_contoh}. The frozen magnon method itself is based on the collective transverse fluctuations of magnetization, which are in the opposite direction to the Stoner excitations (longitudinal spin fluctuations). To calculate the spin stiffness, the frozen magnon method is more efficient than the real-space method proposed by Liechtenstein et al. \cite{Liech} due to the computational cost, as reported in Ref. \citen{4Padja}. 

The energy of spin waves can be obtained from the deviation of the magnetic moment from its ground state \cite{7.1Jakobson}. To evaluate the energy of spin waves by DFT calculation, we assumed the Heisenberg Hamiltonian, i.e., the total energy obtained from the DFT calculation can be approximated by the energy of the Heisenberg Hamiltonian with the definition of the magnetic moment given in Eq. ({\ref{moment}}). The formulation that relates these two energies can be given as
 \beq
E&=&E\left(M_{i}^{2}\right) -\frac{1}{2N} \sum_{i\neq j}J_{ij}\textit{\textbf{M}}_{i}\cdot\textit{\textbf{M}}_{j}\nonumber\\
     &=& E\left(M_{i}^{2}\right)-\frac{1}{2N} \sum_{i\neq j}J_{ij}M_{i}M_{j}\left\{\cos\left[\textit{\textbf{q}}\cdot \left(\textit{\textbf{R}}_{i}-\textit{\textbf{R}}_{j}\right)\right]\right.\nonumber\\
		& &\left.\times\sin\theta_{i}\sin\theta_{j}+\cos\theta_{i}\cos\theta_{j}\right\},
		\label{Heisenberg} 
\eeq
where $i\neq j$ is imposed to avoid the double counting of sites and $N$ denotes the number of unit cells. Here, $E$ and $E(M_{i}^{2})$ are the total energy and ground-state energy for  the unperturbed system, respectively. For convenience, we apply the Fourier transformation for the exchange coupling constant $J_{ij}$  
\beq
J_{\textit{\textbf{q}}}=-\frac{1}{N}\sum_{i\neq j}J_{ij}e^{i\textit{\textbf{q}}\cdot\left(\textit{\textbf{R}}_{i}-\textit{\textbf{R}}_{j}\right)}.\label{exchange}
\eeq
For a small cone angle $\theta$, after substituting Eq. (\ref{exchange}) into Eq. (\ref{Heisenberg}), we obtain
 \beq
\textrm{Re}J_{\textit{\textbf{q}}}-J_{\mathbf{0}}=\left[\frac{1}{M^{2}}\frac{\partial^{2}E}{\partial\theta^{2}}\right]_{\theta=0}.\label{diff}
\eeq
Hereafter, the above equation will be used in formulating the magnon dispersion relation. To derive the magnon dispersion relation, we change the classical Heisenberg Hamiltonian $H$ to the quantum version and use the latter to describe the dynamics of the magnetic moment as
\beq
i\hbar\frac{d\hat{\textit{\textbf{M}}}_{i}}{dt}=\left[\hat{\textit{\textbf{M}}}_{i},\hat{H}\right].\label{dynmom}
\eeq 
In addition, the operator of the lattice magnetic moment always obeys the commutation relations
\beq
\left[\hat{\textit{\textbf{M}}}_{i}^{\alpha},\hat{\textit{\textbf{M}}}_{j}^{\beta}\right]=i\mu_{B}\delta_{ij}\epsilon_{\alpha\beta\gamma}\hat{\textit{\textbf{M}}}_{i}^{\gamma},\label{commut}
\eeq
where the lattice sites and Cartesian coordinates are described by Roman and Greek indexes, respectively. The expression in Eq. (\ref{commut}) depends on the convention, for comparison, one can refer to Refs. \citen{3Halilov} and \citen{16Gebauer}. Moreover, observing that $\hat{\textit{\textbf{M}}}_{i}^{2}$ commutes with all $\hat{\textit{\textbf{M}}}_{i}$, the magnitude of the magnetic moment should not depend on time, $dM_{i}/dt=0$. 

\indent Since the magnetic moment vectors now depend on time, we should take into account the time variable that describes the rotation around the spin rotation axis. We can see directly in the formulation of the magnetic moment in Eq. (\ref{moment}) that we should insert a constant frequency $\omega_{\textit{\textbf{q}}}$ into the angle $\varphi$ to obtain $\varphi_{i}(t)=\textit{\textbf{q}}\cdot \textit{\textbf{R}}_{i}+\omega_{\textit{\textbf{q}}} t$. Substituting Eq. (\ref{moment}) into Eqs. (\ref{dynmom}) and (\ref{commut}), we derive the relation
 \beq
\hbar\omega_{\textit{\textbf{q}}}=2\mu_{B}M\left(\textrm{Re}J_{\textit{\textbf{q}}}-J_{\mathbf{0}}\right).\label{relation}
\eeq 
Comparing Eq. (\ref{diff}) with Eq. (\ref{relation}), we directly conclude that the magnon dispersion relation can be given as
\beq
\hbar\omega_{\textit{\textbf{q}}}=\left[\frac{2\mu_{B}}{M}\frac{\partial^{2}E}{\partial\theta^{2}}\right]_{\theta=0}.\label{magnon}
\eeq
As immediately observed from Eq. (\ref{magnon}), we should calculate $E(\theta)$ for several $\theta$ and evaluate the second derivative of $E(\theta)$ by using the available methods, such as the finite difference method as suggested by Gebauer \cite{16Gebauer}. However, this method is not so efficient since it includes several computations before the calculation of the magnon dispersion relation. There is another way, namely, replacing the second derivative $\partial^{2}E/\partial\theta^{2}$ with the analytical form in terms of the total energy difference $\Delta E(\textit{\textbf{q}},\theta)=E(\textit{\textbf{q}},\theta)-E(\mathbf{0},\theta)$. We therefore rewrite Eq. (\ref{magnon}) as  
\beq
\hbar\omega_{\textit{\textbf{q}}}=\lim_{\theta\rightarrow 0} \frac{4\mu_{B}}{M}\frac{\Delta E(\textit{\textbf{q}},\theta)}{\sin^{2}\theta}.\label{magnonf}
\eeq    
This relation was derived by K$\ddot{\textrm{u}}$bler and applied to calculate the spin stiffness for 3$d$ transition metals \cite{7Kubler}. It can derive Eq. (\ref{magnon}) by applying L'H$\hat{\textrm{o}}$pital's rule. The other methods to derive the magnon dispersion are given in Refs. \citen{16.1Antropov, 16.2Antropov, 16.3Savrasov}. 

\indent The schematic procedure to obtain the curve of the magnon dispersion relation to obtain the spin stiffness by using an LCPAO is as follows. First, we choose successive spiral vectors $\textit{\textbf{q}}$ near the $\Gamma$ point; in this paper we select the range [0, 1] {\AA}$^{-1}$ for bcc-Fe and set the cone angle $\theta$. Then, based on the selected $\theta$, we calculate the total energy difference between $\textit{\textbf{q}}=0$ and each of the chosen higher $\textit{\textbf{q}}$, and add the right-hand side of Eq. (\ref{magnonf}). Finally, we evaluate the left-hand side of Eq. (\ref{magnonf}) with $\hbar\omega_{\textit{\textbf{q}}}=Dq^{2}(1-\beta q^{2})$ to compute the spin stiffness $D$. The above procedure has been employed by several authors \cite{2Rosengaard,3Halilov,7Kubler} to calculate the spin stiffness for bcc-Fe, which will be compared with our result in Sec. 3. For the graphene nanoribbon, we follow a somewhat different approach, in which sufficiently lower magnitudes of spiral vectors $\textit{\textbf{q}}$ than those for the case of bcc-Fe case are set to fix the magnitudes of the magnetic moment as described later in Sect. 3.
    
\section{Results and Discussion}    
\subsection{Carrier-induced spin-spiral ground state in one-dimensional hydrogen chain}
\begin{figure}[h!]
\vspace{1mm}
\centering
\includegraphics[scale=0.55, width =!, height =!]{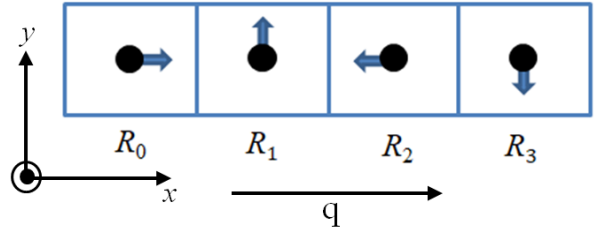}
\vspace{-1 mm}
\caption{\label{H-chain} (Color online) Simple illustration of the one-dimensional hydrogen chain along the $x$ axis for the case of $\textit{\textbf{q}}=0.25$, which corresponds to $90^{\circ}$ rotation, with the initial angles $(\theta_{0},\varphi_{0})=(90^{\circ},0^{\circ})$. These angles represent the direction of the magnetic moment. Here, $\textit{\textbf{q}}$ is defined in units of $2\pi/a$ ($a$ is the lattice constant).} 
\end{figure}

\indent By applying the MFA to the one-dimensional single-orbital Hubbard model, Inoue and Maekawa have shown that a magnetic phase transition takes place from antiferromagnetic (AFM) state to a spiral state at $T=0$ in a perovskite manganite by introducing hole doping \cite{17Inoue}. Another carrier-induced magnetic phase transition in a perovskite manganite has also been reported by applying the constraint scheme method within noncollinear DFT \cite{sawada2}. Since the perovskite manganite has one electron at the $e_{g}$ state of the Mn site, it can be modeled by a one-dimensional hydrogen chain, similarity to the one-dimensional single-orbital Hubbard model. Therefore, we expect that the magnetic phase transition takes place in a one-dimensional hydrogen chain. We used one valence orbital $s$ as a basis set specified by H7.0-$s1$, which indicates that the H atom has a 7.0 Bohr cutoff radius. The direction of the spiral vector $\textit{\textbf{q}}$ was selected to be along the $x$ axis with a $30 \times 1 \times 1$ $k$ point grid and a cutoff energy of 150 Ryd as illustrated in Fig. \ref{H-chain}. 
We also employed 2.5 {\AA} for the lattice constant and used the generalized gradient approximation (GGA) \cite{18Perdew} for the exchange-correlation potential.
\begin{figure}[h!]
\vspace{-8 mm}
\quad\quad\quad\quad\quad\quad\quad\quad\quad\includegraphics[scale=0.6, width =!, height =!]{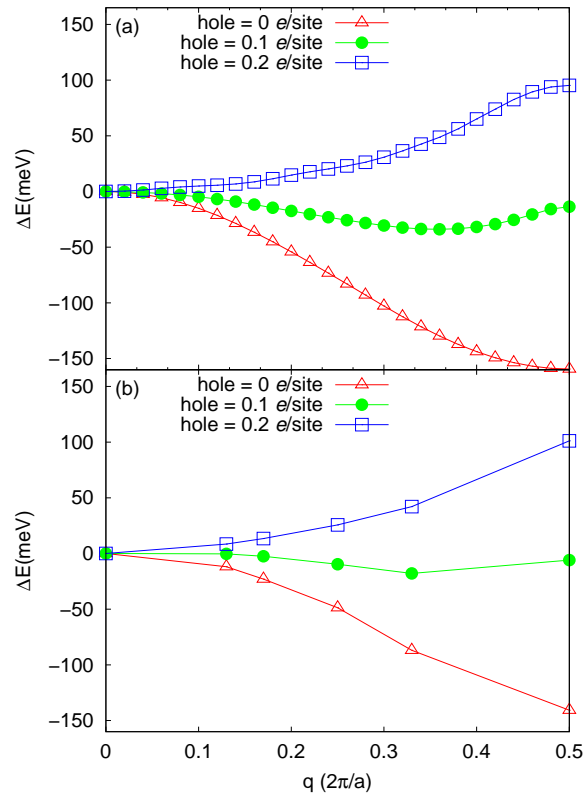}
\vspace{-20mm}
\caption{\label{spiral_H_supercell}(Color online) Hole doping dependence of spiral vector vs total energy difference for the one-dimensional hydrogen chain using the GBT (a) and the supercell calculation (b). Here, the total energy difference $\Delta E$ is obtained by subtracting the total energy of each \textit{\textbf{q}} from that for each hole doping with respect to \textit{\textbf{q}} $=0$.}
 \vspace{-6mm}
\end{figure}
\begin{figure}[h!]
\vspace{2mm}

\quad\quad\quad\quad\quad\quad\quad\quad\quad\quad\quad\includegraphics[scale=0.6, width =!, height =!]{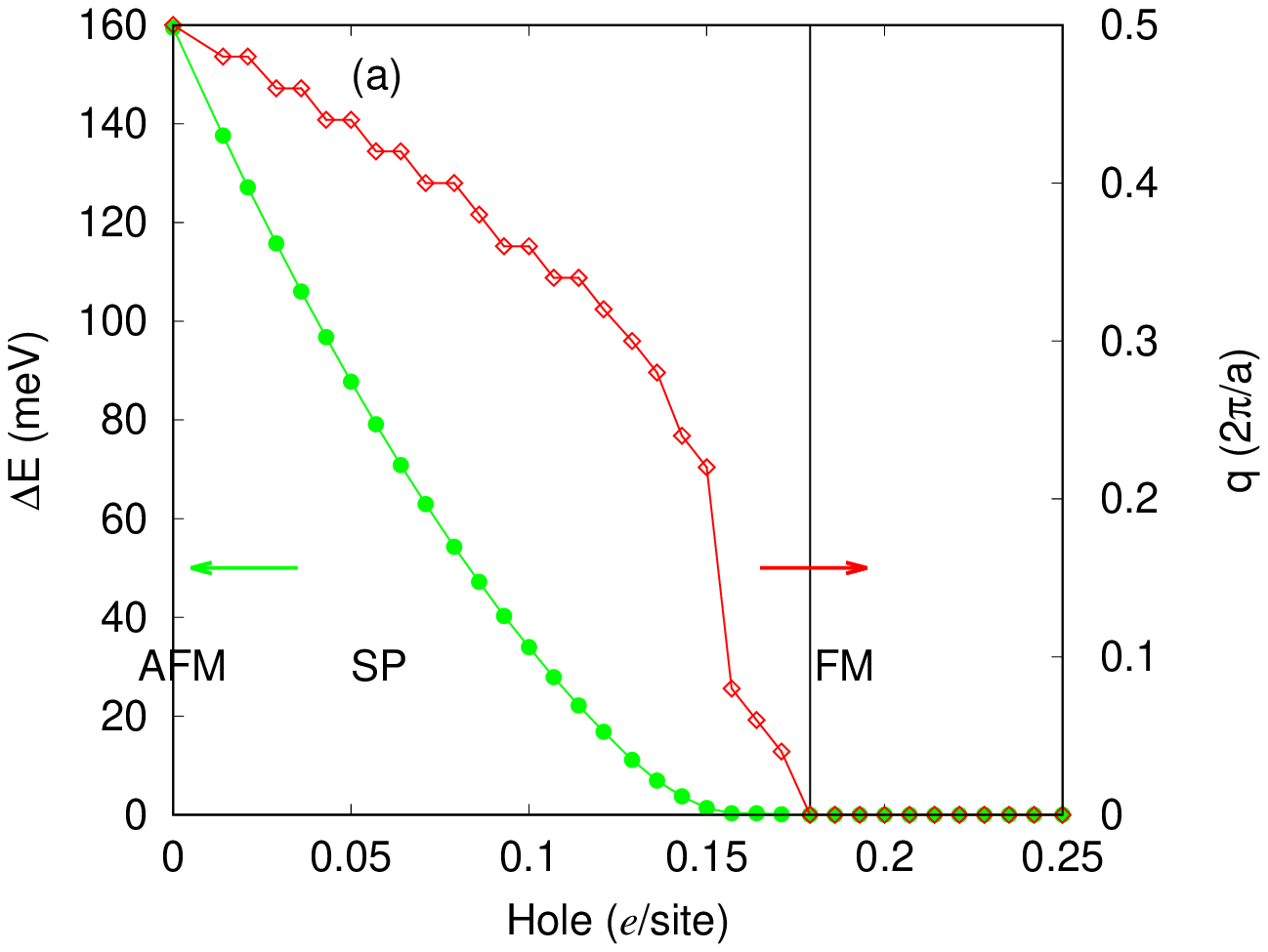} 
\begin{center}
\quad\quad\quad\quad\quad\includegraphics[scale=0.6, width =!, height =!]{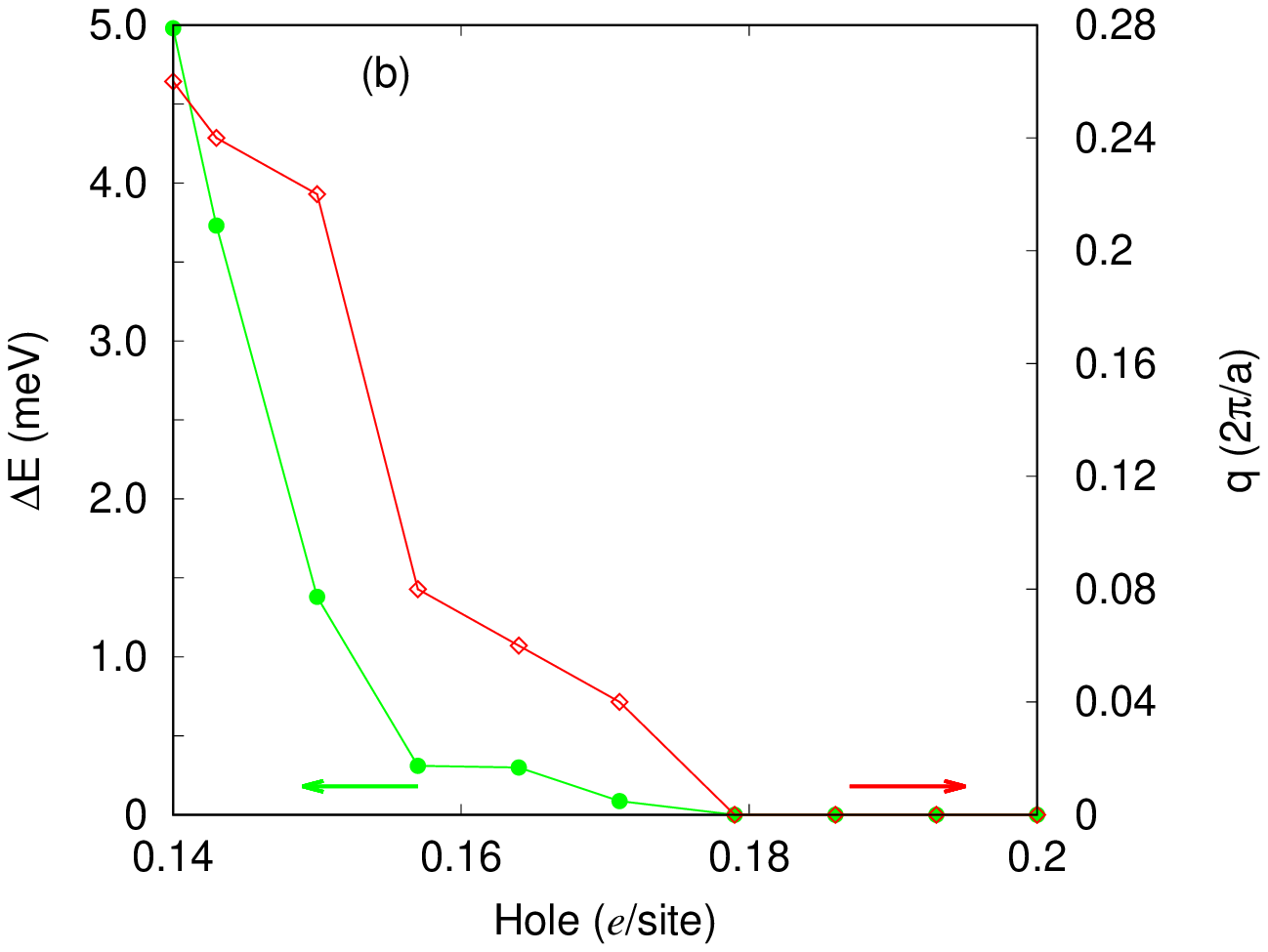}
\end{center}
\vspace{5mm}
\caption{ (Color online) (a) Hole doping dependences of the total energy difference $\Delta E$ and the spiral vector $\textit{\textbf{q}}$ for the one-dimensional hydrogen chain using the GBT. The total energy difference at each hole doping is defined as $E(0)-E(\textit{\textbf{q}})$, where $E(\textit{\textbf{q}})$ is the total energy of the ground state with spiral vector $\textit{\textbf{q}}$. (b) Plot around phase boundary between spiral and FM ground states.} \label{phase} 
\vspace{-4mm}
\end{figure}

Figure {\ref{spiral_H_supercell}}(a) shows the $\textit{\textbf{q}}$-dependent total energy difference for nondoping, 0.1 $e$/site hole doping, and 0.2 $e$/site hole doping for a one-dimensional hydrogen chain. The calculated ground states for nondoping, 0.1 $e$/site hole doping, and 0.2 $e$/site hole doping are the AFM state at $q=0.5$, the spiral (SP) state at $q=0.36$, and the ferromagnetic (FM) state at $q=0$, respectively. For comparison, we provide the supercell calculation for several cells in Fig. {\ref{spiral_H_supercell}}(b). Here, we only consider 2, 3, 4, 6, and 8 supercells from right to left, which are associated with rotation angles of 180$^{\circ}$, 120$^{\circ}$, 90$^{\circ}$, 60$^{\circ}$, and 45$^{\circ}$, respectively. By comparing the two figures, we observe the same profiles, as mentioned previously, as those when using the GBT, and the total energy differences for all hole doping are of the same order. As shown in Fig. {\ref{phase}}, amounts of detailed calculations, including those for very small hole doping in the range of 0 $\leq$ hole $\leq$ 0.2 $e$/site, are performed to explore three separate ranges, i.e., the nondoping case is the AFM ground state, 0 $<$ hole $\leq$ 0.171 $e$/site is the SP ground state, and hole $>$ 0.171 $e$/site is the FM ground state. This result is similar to the previous result obtained by Inoue and Maekawa using the one-dimensional single-orbital Hubbard model \cite{17Inoue}.

Note that the magnetic phase transition relies on the competition between the kinetic energy and exchange interaction, which depends on the lattice constant in our model. We use the lattice constant of 2.5 {\AA} as an example to investigate this competition. We have examined the lattice constant dependence of the magnetic ground state. For example, when we increase the lattice constant to 3.0 {\AA}, the ground state for the nondoping case is still the AFM state. On the other hand, in the case of 0.1 $e$/site hole doping, the ground state changes from the SP state to the FM state. It seems that varying the lattice constant tunes the new magnetic phase transition due to the competition between the kinetic energy and the exchange interaction. 

We used the minimal basis set, which contains only an $s$ orbital, since our purpose is to test the implementation of the GBT by reproducing the one-dimensional single-orbital Hubbard model. Here we comment on the basis set dependence of our model. We investigate the inclusion of the $p$ orbital, i.e., $s2p1$ orbitals, in the hydrogen chain system to evaluate the effect. For simplicity, we only consider the nondoping case. When $s2p1$ orbitals are applied, the AFM state is no longer the ground state when using both the GBT and the supercell calculation. We argue that PAO cutoff radius of 7.0 Bohr (3.7 {\AA}) used here is too large compared with the lattice constant of 2.5 {\AA} when including the $p$ orbital. This means that the wavefunction using $s2p1$ orbitals for the H atom will cover up the next nearest-neighbor of the H atom much more than that when using the $s$ orbital. This situation is different from that of the one-dimensional single-orbital Hubbard model with the nearest-neighbor approximation. Therefore, if we want to obtain the same ground state using the $s$ orbital and $s2p1$ orbitals, we should increase the lattice constant or decrease the cutoff radius so that the effect of the $p$ orbital can be decreased. A clear description of the relationship between the cutoff radius and the number of orbitals will be given when considering the case of bcc-Fe in Sect. 3.2.

\subsection{Magnon dispersion relation of bcc-Fe}
\begin{figure}[h!]
\vspace{2 mm}
\centering
\includegraphics[scale=0.8, width =!, height =!]{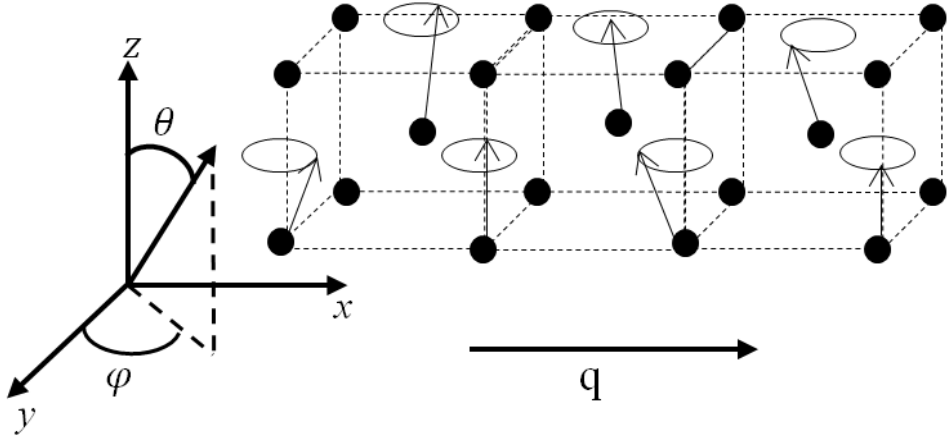}
\vspace{0.01cm}
\caption{\label{bcc} Conical spin spiral for bcc-Fe along the $x$ direction with a small deviation from the FM ground state. The initial magnetic moment is represented by the initial angles $(\theta_{0},\varphi_{0})=(10^{\circ},0^{\circ})$.}
 \vspace{1 mm}
\end{figure}

\indent We now focus on how to calculate the spin stiffness for bcc-Fe using an LCPAO. The main reason why we chose bcc-Fe is that the spin stiffness has been well provided theoretically and experimentally as shown in Table \ref{table}. In performing the first-principles calculation, we used the local spin density approximation (LSDA) \cite{19Ceperley}, we set the cutoff energy to 300 Ryd, and varied both the cutoff radius and the number of orbitals. In this paper, as shown in Fig. \ref{bcc}, we fixed the cone angle to $10^{\circ}$ by using the constraint DFT provided in the OPENMX code. Note that the total energy difference cannot be fitted by a fourth-order fit if using a small number of $k$ points. This is due to the insufficient $k$ point mesh for the metallic system. Therefore, we used a $50 \times 50 \times 50$ $k$ point mesh to overcome this problem.
For the next discussion, we define the notation of basis sets Fe4.0-$s2p2d1$, where Fe is the atomic symbol for iron, 4.0 is the cutoff radius in units of Bohr, and $s2p2d1$ indicates that two primitive orbitals are employed for $s$ and $p$ orbitals and one primitive orbital is employed for the $d$ orbital.
\begin{figure*}[h!]
\vspace{4mm}
\quad\quad\quad\includegraphics[scale=0.6, width =!, height =!]{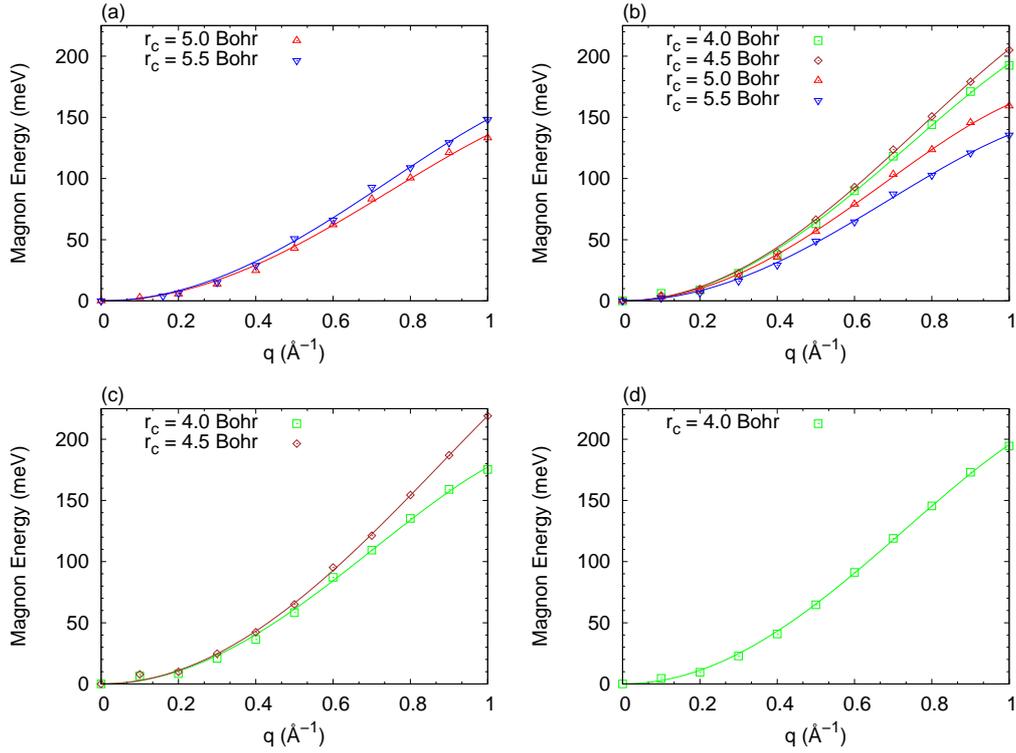}
\vspace{-52mm}
\caption{\label{magnon_disp} (Color online) Magnon energy of bcc-Fe near $\textit{\textbf{q}}=0$ for several cutoff radii and basis sets (a) $s2p2d1$, (b) $s2p2d2$, (c) $s3p3d3$, (d) $s3p3d3f2$. Here, all the solid lines represent the fitting function $\hbar\omega_{\textit{\textbf{q}}}=Dq^{2}(1-\beta q^{2})$.} 
\end{figure*}

\begin{figure}[h!]
\vspace{4mm}
\centering\includegraphics[scale=0.4, width =!, height =!]{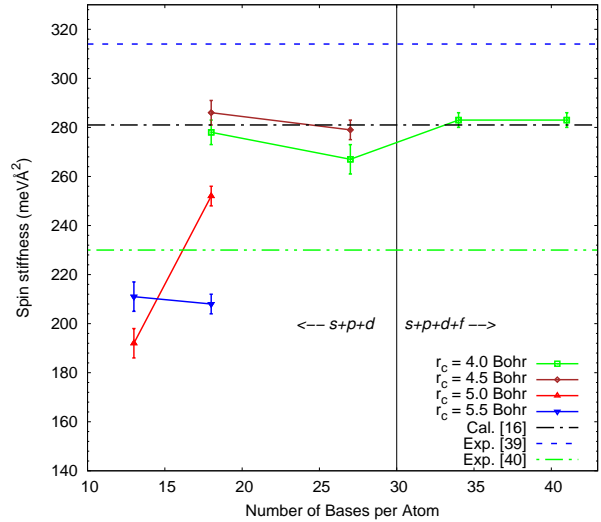}
\vspace{8mm}
\caption{\label{conv}(Color online) Basis-set dependence of spin stiffness for bcc-Fe. The reported calculations and experimental data are also shown.} 
\vspace{-6 mm}
\end{figure}

\indent The accuracy of the wavefunction is determined by the cutoff radius and the number of orbitals. First, if we use a short cutoff radius, the number of orbitals should be sufficiently large, otherwise the insufficient number of orbitals will result in a poor fitting, i.e., deviation between the data and fitting line. Second, if we use a long cutoff radius, a small number of orbitals should be used to avoid a poor fitting due to overcompleteness. Figure {\ref{magnon_disp}} provides a good fitting for the relationship between the cutoff radius and the number of orbitals. Owing to the same value of spin stiffness when using $s3p3d3f2$ orbitals, the magnon energy using $s3p3d3f1$ orbitals is not shown. Here, we can see that the cutoff radius of 4.0 Bohr is sufficiently short since the number of orbitals can be increased to a large number. On the other hand, it is also shown that the cutoff radii of 5.0 and 5.5 are sufficiently long to increase the number of orbitals over $s$2$p$2$d$2 orbitals due to overcompleteness. We summarize our calculated spin stiffness and report the theoretical and experimental spin stiffnesses in Table \ref{table} and Fig. {\ref{conv}}. As observed in Fig. {\ref{conv}}, we also confirm that our LCPAO result is in good agreement with the result using the plane-wave basis set and pseudopotential calculated by Gebauer \cite{16Gebauer}.   
\begin{table}[ht]
\vspace{-4.5 mm}
\caption{Calculated spin stiffness for bcc-Fe and comparison with other calculations and experimental results.}   
\centering 
\begin{tabular}{c c } 
\hline 
$D_{cal}$ (meV{\AA}$^{2}$)& $D_{exp}$ (meV{\AA}$^{2}$)\\ 
\hline 
283, 247 \cite{2Rosengaard}, 250 \cite{4Padja}, &314 \cite{20Stringfellow}, 230 \cite{21Lynn}\\[1ex] 
 281 \cite{16Gebauer}, 355 \cite{7Kubler}, 313 \cite{Shallcross}& 280 \cite{21aPauthenet}, 307 \cite{22Loong}\\ 
\hline 
\end{tabular}
\label{table} 
\vspace{-5 mm}
\end{table}
Our claim is that the reliable spin stiffness for bcc-Fe is about 283 meV{\AA}$^{2}$, obtained using a 4.0 Bohr cutoff radius, and its convergence can be easily achieved. We also state that the fitting error obtained by our spin stiffness calculation is 3 meV{\AA}$^{2}$ for Fe4.0-$s3p3d3f2$, which is acceptable considering the experimental result of about 15 meV{\AA}$^{2}$ \cite{22Loong}.              

\indent From the above results, we should carefully examine the relationship between the number of orbitals and the cutoff radius to obtain reliable results. A similar dependence on these parameters was also reported in a recent paper \cite {Yoon} on implementing the magnetic-force theorem using an LCPAO for a metallic system such as bcc-Fe. In Ref. \citen {Yoon}, it was stated that the appropriate choice of parameters, such as the cutoff radius and the number of orbitals, can give more reliable results compared with the experimental results.     
\subsection{Carrier-induced change in spin stiffness for graphene nanoribbon} 
\indent Graphene nanoribbons have been reported to be promising materials for use in spintronic applications \cite{nanorib1,nanorib2}. In graphene nanoribbons, magnetism can be generated by various treatments, such as applying an external electric field \cite{30nanorib1, 31nanorib2, 32nanorib3} and arranging the magnetic moment at different edges \cite{33nanorib4, 34nanorib5, 35nanorib6}.  

\indent In the case that weak ferromagnetism appears upon the doping of electrons or holes, it is of interest to explore another physical property in depth, i.e., the change in spin stiffness with respect to the carrier doping. As predicted in a previous paper \cite{36nanorib7}, zigzag graphene nanoribbons (ZGNRs) should have higher spin stiffness than 3$d$ transition elements due to the very small magnetic moment of C atoms of about 0.28 $\mu_{\textrm{B}}$ at the two edges. By performing a supercell calculation, Yazyev and Katsnelson have obtained a spin stiffness of $D=$2100 meV{\AA}$^{2}$ for a ZGNR \cite{37nanorib8}. Our motivation for using the GBT is to compute the spin stiffness more accurately by using the total energy difference of long-wavelength spin waves. We also investigated the carrier dependence of spin stiffness. A major advantage of using the GBT instead of performing a supercell calculation is that the computational cost is lower if considering long-wavelength spin waves as performed in this paper.   
\begin{figure}[h!]
\vspace{6mm}
\begin{center}
\includegraphics[scale=0.55, width =!, height =!]{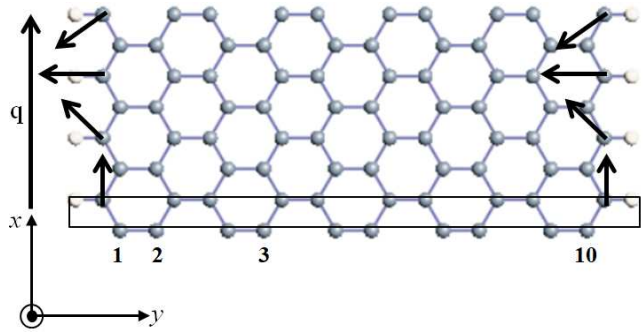}
\end{center}
\vspace{0.01cm}
\caption{\label{graphene_model}(Color online) Flat spiral configurations in a ferromagnetic ZGNR model with initial angles $(\theta_{0},\varphi_{0})=(90^{\circ},0^{\circ})$. The unit cell is represented by the black rectangle while the spiral vector is parallel to the periodic direction along the $\emph{x}$ axis. Here, C and H atoms are represented by blue and white spheres, respectively.} 
\vspace{-2 mm}
\end{figure}
                 
\indent We constructed a flat spiral ($\theta=90^{\circ}$) by using a ferromagnetic edge state unit cell with 10 ZGNRs, which means that the width of the ribbon is 10, as shown in Fig. {\ref{graphene_model}}, by imposing a constraint on the spin orientation. A structural optimization was conducted using a nonmagnetic structure, and the lattice constant of graphite (2.46 {\AA}) obtained from the experiment was set as the length of the unit cell along the \emph{x} axis (periodic direction). A cutoff energy of 150 Ryd was set to obtain convincing results. Here, we used C4.0-$s2p2$ and H6.0-$s2p1$ for the basis sets, meaning that two valence orbitals for the $s$ and $p$ orbitals are assigned to C atoms while two valence orbitals for $s$ and one valence orbital for $p$ are assigned to H atoms. The cutoff radii for C and H atoms are 4.0 and 6.0 Bohr, respectively. Using these assignments, the self-consistency of all calculations can be confirmed with the magnetic moment of 0.28 $\mu_{\textrm{B}}$ of C atoms at the two edges. Following previous studies \cite{33nanorib4, 34nanorib5}, we chose the GGA \cite{18Perdew} as the exchange-correlation potential, set a $90 \times 1 \times 1$ \emph{k} point mesh, and applied the length of the vacuum area (nonperiodic directions) larger than 25 {\AA} to ensure the vacuum condition.             

\begin{figure}[h!]
\vspace{-10mm}
\quad\quad\includegraphics[scale= 1.1, width = !, height =!]{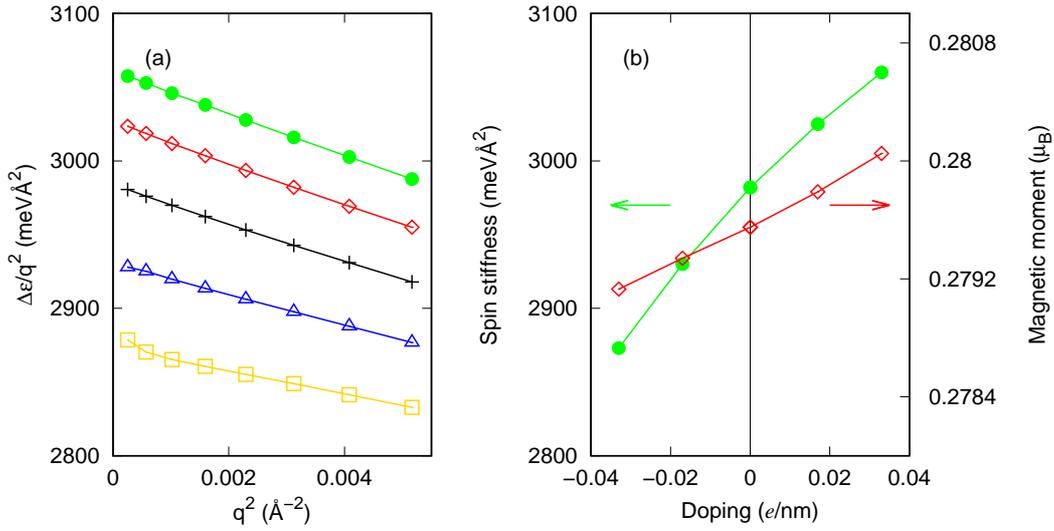}
\vspace{-80mm}
\caption{(Color online) (a) Quadratic spiral vector $\textit{\textbf{q}}^{2}$ dependence of total energy difference per total magnetic moment of C atom at the edges $\Delta \varepsilon=2\Delta E/M$ divided by $\textit{\textbf{q}}^{2}$ for 10 ZGNR. Plus signs, diamonds, filled circles, triangles, and empty squares denote nondoping and doping of 0.017, 0.033, -0.017, and -0.033 $e$/nm, respectively. (b) Doping dependence of spin stiffness and magnetic moment at $\textit{\textbf{q}}=0$. Filled circles and diamonds represent the spin stiffness and magnetic moment, respectively.} 
\label{qkuadrat}
\vspace{-6 mm}
\end{figure}

\indent To use the frozen magnon method, the magnitude of the magnetic moment should be constant. In the case of a ZGNR, the magnitude of the magnetic moment is sensitive to the spiral vector $\textit{\textbf{q}}$. Therefore, we used the spiral vectors $\textit{\textbf{q}}$, whose the magnitudes are sufficiently lower than those used for bcc-Fe. After evaluating the total energy difference $\Delta E=E(\textit{\textbf{q}})-E(0)$ for each $\textit{\textbf{q}}$, we fit the total energy difference to obtain $\kappa$ using the fourth-order dispersion $\Delta E=\kappa q^{2}(1-\beta q^{2})$ instead of the quadratic dispersion used by Yazyev and Katsnelson \cite{37nanorib8}. This was to obtain the dependence of the spin stiffness on the quadratic spiral vectors, see Fig. {\ref{qkuadrat}}(a), which can only be obtained by fourth-order fitting. In this calculation, we applied the following values of doping ($e$/nm): -0.033, -0.017, 0, 0.017, and 0.033. Spin stiffness constants were determined using the expression $D=2\kappa/M$, where $M$ denotes the total magnetic moment of C atoms at the two different edges.
\begin{table}
\vspace{2 mm}
\caption{Calculated spin stiffness $D$ (meV{\AA}$^{2}$) for ferromagnetic edge state in ZGNR for several values of doping. For the nondoping case, 2100 meV{\AA}$^{2}$ is the spin stiffness obtained by Yazyev and Katsnelson by supercell calculation \cite{37nanorib8}.}   
\centering 
\begin{tabular}{c c c c c c} 
\hline 
Doping ($e$/nm)& -0.033& -0.017& 0.0& 0.017& 0.033\\ 
\hline 
$D$ (meV{\AA}$^{2}$)& 2873& 2930& 2982,& 3025&3060\\ 
& & & 2100 \cite{37nanorib8}& &\\
\hline 
\end{tabular}
\vspace{0.01cm}
\label{stifness_graphene} 
\vspace{-6 mm}
\end{table}

\indent The evaluated spin stiffness constants $D$ are provided in Table {\ref{stifness_graphene}} and Fig. {\ref{qkuadrat}} together with the previous result obtained by Yazyev and Katsnelson \cite{37nanorib8}. As seen in Table {\ref{stifness_graphene}}, we obtained a spin stiffness of 2982 meV{\AA}$^{2}$ for the nondoping case. Even though this value is larger than that of Ref. \citen{37nanorib8}, the order is the same. Next, we discuss the change in spin stiffness by resulting  from carrier doping. We initially examine the convergence of each value of spin stiffness. As immediately observed in Fig. {\ref{qkuadrat}}(a), good convergence of the spin stiffness for each quadratic spiral vector was achieved. This means that the trends show a suitable change in the spin stiffness using this set of spiral vectors. The change in spin stiffness together with the appropriate magnetic moment is given in Fig. {\ref{qkuadrat}}(b). We found a difference between the carrier doping dependence of spin stiffness for hole and electron doping. As hole doping increases, the spin stiffness increases, while the spin stiffness decreases as electron doping increases. In addition, we also observed the same tendency of the magnetic moment at $\textit{\textbf{q}}=0$ with respect to carrier doping as also shown in Fig. {\ref{qkuadrat}}(b). Since the excitations of spin waves are governed by the rotation of magnetic moments of magnetic atoms, we justify that the resulting tendency is driven by the magnitude of the magnetic moment. This justification is supported by considering the perspective of the classical Heisenberg model, in which the total energy difference in our system is proportional to the magnitude of the magnetic moments $M$ of the two C atoms at the two edges where the exchange coupling constant $J_{ij}$ is fixed. Since the spin stiffness is proportional to the total energy difference, the tendency of spin stiffness follows the tendency of the total energy difference. From all the above observations, it seems that the spin stiffness can be tuned by varying the carrier doping.                      

We have verified the basis set dependence of the results. Since the ZGNR is a $p$-electron system, the minimal basis set should contain $s$ and $p$ orbitals. We also examine the inclusion of $d$ orbital for C atoms to observe the tendency of spin stiffness. We increase the number of orbitals for C atoms up to $s$2$p$2$d$1 orbitals. For the nondoping case, the spin stiffness for $s$2$p$2$d$1 orbitals is 3036 meV{\AA}$^{2}$, larger than the spin stiffness of 2982 meV{\AA}$^{2}$ for the case of $s$2$p$2 orbitals, a difference of about 2\%. This means that the use of these two different orbitals will not significantly affect the rigidity of spin stiffness. Therefore, by comparing $s$2$p$2 and $s$2$p$2$d$1 orbitals, we clarify that the use of orbitals up to $p$ orbitals is sufficient to calculate the spin stiffness.

It is interesting to note that if we increase the width of the ribbons, the spin stiffness is enhanced, as shown in Table \ref{N_ZGNR}. For a fixed lattice constant, it seems that the distance between the two edges of C atoms contributes to the total energy difference. This means that we need much more energy to excite spin waves. This is also supported by the enhancement of the magnetic moment with increasing width of the ribbons, as shown in Table \ref{N_ZGNR}.
\begin{table}
\vspace{3 mm}
\caption{Calculated spin stiffness as a function of ribbon width for the nondoped ZGNR.}   
\centering 
\begin{tabular}{c c c} 
\hline 
Width& $D$ (meV{\AA}$^{2}$)&Moment per C atom ($\mu_{\textrm{B}})$\\ 
\hline 
8& 2775&0.27374\\ 
10& 2982&0.27955\\ 
12& 3219&0.28302\\ 
\hline 
\end{tabular}
\vspace{0.01cm}
\label{N_ZGNR} 
\vspace{-6 mm}
\end{table}

\indent For further discussion, we attempt to find the likelihood of the appearance of a carrier-induced spin-spiral ground state, as conducted previously for a one-dimensional hydrogen chain. We increase the concentration of both hole and electron doping to 2.85 $e$/nm as in Ref. \citen{34nanorib5}. However, we do not find a spiral ground state despite the increased doping. At 2.85 $e$/nm, the nonmagnetic (NM) state appears, similar to Ref. \citen{34nanorib5}. Here, we give some comments on the use of the GBT in the case of the ZGNR. In fact, a different version of the GBT can be applied in the case of the ZGNR, which can be found in Ref. \citen{38GBTnano}. In this reference, the authors implemented the GBT for a ZGNR with a collinear magnetic structure to study the half-metallic property. Even though the concept of the GBT is similar to ours, they used the rotation symmetry to induce bending of the ZGNR.
\section{Conclusions}

\indent The GBT has been successfully implemented in the OPENMX code in order to perform a spin spiral calculation to replace the supercell calculation if SOC is neglected. As directly seen, by using an LCPAO and norm-conserving pseudopotentials in the OPENMX code, we can obtain reliable results compared with other methods.

\indent We have proven that the spiral ground state can be induced by increasing hole doping for the case of a one-dimensional hydrogen chain. In this case, different levels of hole doping induce three states, i.e., antiferromagnetic (AFM) - spiral (SP) - ferromagnetic (FM) ground states. It is also seen that the results obtained by supercell calculation are in good agreement with those obtained using the GBT.   

\indent For bcc-Fe, one must be careful to determine the number of orbitals used as basis sets to ensure the quality of the fitting in the calculation. In this case, we found that increasing the number of orbitals to obtain a good result can only be done for a short cutoff radius, which has been verified in Ref. \citen{13Ozaki}, such as for determining the equilibrium lattice constant. Therefore, although we can also apply a long cutoff radius, it is recommended that the minimal number of orbitals is used as the basis sets. Conversely, by using a short cutoff radius, we should use a sufficiently large number of orbitals to attain reliable results. 

\indent For the ZGNR, our calculated spin stiffness using the GBT has the same order as that obtained previously by Yazyev and Katsnelson \cite{37nanorib8} by supercell calculation. We show that the spin stiffness can be controlled by changing the carrier doping; increasing the hole or electron doping increases or decreases the spin stiffness. In addition, it is also shown that the magnitude of the magnetic moment behaves similarly to the spin stiffness as the carrier doping changes. 

From the results in this paper, we conclude that the implementation of the GBT using an LCPAO is successful and can be applied to more complex magnetic materials.   
                     
\section*{Acknowledgments}
This work was supported by Japan Society for the Promotion of Science (JSPS) Grants-in-Aid for Scientific Research on Innovative Area, "Nano Spin Conversion Science" (Grant Nos. 15H01015 and 17H05180). It was also supported by a JSPS Grant-in-Aid for Scientific Research on Innovative Area, "Discrete Geometric Analysis for Material Design" (Grant No. 18H04481). It was partially supported by a JSPS Grant-in-Aid on Scientific Research (Grant No. 16K04875). All computational calculations were carried out using ISSP supercomputers located at the University of Tokyo. One of the authors (T.B.P.) would like to acknowledge the Directorate General of Higher Education (DIKTI), Indonesia, for the scholarship program.


\begin{thebibliography}{99}
\bibitem{0Coey} J. M. D. Coey, Can. J. Phys. \textbf{65}, 1210 (1987).
\bibitem{1Sandratskii} L. M. Sandratskii, Adv. Phys. \textbf{47}, 91 (1998).
\bibitem{2Rosengaard} N. M. Rosengaard and B. Johansson, Phys. Rev. B \textbf{55}, 14975 (1997). 
\bibitem{3Halilov} S. V. Halilov, H. Eschrig, A. Y. Perlov, and P. M. Oppeneer, Phys. Rev. B \textbf{58}, 293 (1998). 
\bibitem{4Padja} M. Pajda, J. Kudrnovsk$\acute{\textrm{y}}$, I. Turek, V. Drchal, and P. Bruno, Phys. Rev. B \textbf{64}, 174402 (2001).
\bibitem{7.1Jakobson} A. Jakobsson, B. Sanyal, M. Le$\breve{\textrm{z}}$a{\'{i}}c, and S. Bl$\ddot{\textrm{u}}$gel, Phys. Rev. B \textbf{88}, 134427 (2013).
\bibitem{7.2Jakobson} A. Jakobsson, P. Mavropoulos, E. $\c{S}$a$\textrm{\c{s}}$io$\breve{\textrm{g}}$lu, S. Bl$\ddot{\textrm{u}}$gel, M. Le$\breve{\textrm{z}}$a{\'{i}}c, B. Sanyal, and I. Galanakis, Phys. Rev. B \textbf{91}, 174439 (2015). 
\bibitem{5Heide} M. Heide, G. Bihlmayer, and S. Bl$\ddot{\textrm{u}}$gel, Physica B \textbf{404}, 2678 (2009).
\bibitem{6Openmx} T. Ozaki, H. Kino, J. Yu, M. J. Han, N. Kobayashi, M. Ohfuti, F. Ishii, T. Ohwaki, H. Weng, and K. Terakura, Open source package for Material eXplorer (http://www.openmx-square.org).
\bibitem{Yang} W. Yang, Phys. Rev. Lett. \textbf{66}, 1438 (1991).
\bibitem{Ozaki-Terakura} T. Ozaki and K. Terakura, Phys. Rev. B \textbf{64}, 195126 (2001).
\bibitem{Portal}D. S$\acute{\textrm{a}}$nchez-Portal, P. Ordej$\acute{\textrm{o}}$n, E. Artacho, and J. M. Soler, Int. J. Quantum Chem. \textbf{65}, 453 (1997). 
\bibitem{Lewis}J. P. Lewis, P. Ordej$\acute{\textrm{o}}$n, and O. F. Sankey, Phys. Rev. B \textbf{55}, 6880 (1997).
\bibitem{OzakiO} T. Ozaki, Phys. Rev. B \textbf{74}, 245101 (2006).
\bibitem{Ohwaki} T. Ohwaki, M. Otani, T. Ikeshoji, and T. Ozaki, J. Chem. Phys. \textbf{136}, 134101 (2012).
\bibitem{16Gebauer} R. Gebauer, Dr. Thesis, Ecole Normale Sup$\acute{\textrm{e}}$rieure de Lyon, Lyon (1999).
\bibitem{14Gebauer} R. Gebauer and S. Baroni, Phys. Rev. B \textbf{61}, R6459 (2000).
\bibitem{15Kurz} Ph. Kurz, F. F$\ddot{\textrm{o}}$rster, L. Nordstr$\ddot{\textrm{o}}$m, G. Bihlmayer, and S. Bl$\ddot{\textrm{u}}$gel, Phys. Rev. B \textbf{69}, 024415 (2004).
\bibitem{33nanorib4} K. Sawada, F. Ishii, and M. Saito, Appl. Phys. Express \textbf{1}, 064004 (2008).
\bibitem{sawada2} K. Sawada and F. Ishii, J. Phys.: Condens. Matter \textbf{21}, 064246 (2009).
\bibitem{mizuta} Y. P. Mizuta and F. Ishii, Sci. Rep. \textbf{6}, 28076 (2016).
\bibitem{SIESTA} E. Artacho, J. M. Cela, J. Gale, A. Garc{\'{i}}a, J. Junquera, R. M. Martin, P. Ordej$\acute{\textrm{o}}$n, D. S$\acute{\textrm{a}}$nchez-Portal, and J. M. Soler, SIESTA code (https://departments.icmab.es/leem/siesta).
\bibitem{24Garcia} V. M. Garc{\'{i}}a-Su$\acute{\textrm{a}}$rez, C. M. Newman, C. J. Lambert, J. M. Pruneda, and J. Ferrer, Eur. Phys. J. B \textbf{40}, 371 (2004).
\bibitem{HanL1} M. J. Han, T. Ozaki, and J. Yu, Phys. Rev. B \textbf{73}, 045110 (2006).
\bibitem{HanL2} S. Ryee and M. J. Han, Sci. Rep. \textbf{8}, 9559 (2018).
\bibitem{HanL3} S. Ryee and M. J. Han, J. Phys.: Condens. Matter \textbf{30}, 275802 (2018).
\bibitem{13.1Troullier} N. Troullier and J. L. Martins, Phys. Rev. B \textbf{43}, 1993 (1991).
\bibitem{13Ozaki} T. Ozaki and H. Kino, Phys. Rev. B \textbf{69}, 195113 (2004).
\bibitem{13.2Ozaki} T. Ozaki, Phys. Rev. B \textbf{67}, 155108 (2003).
\bibitem{Liech} A. I. Liechtenstein, M. I. Katsnelson, V. P. Antropov, and V. A. Gubanov, J. Magn. Magn. Mater. \textbf{67}, 65 (1987).   
\bibitem{7Kubler} J. K$\ddot{\textrm{u}}$bler, in \emph{Theory of Itinerant Electron Magnetism} (Oxford University Press), Oxford, (2009).
\bibitem{16.1Antropov} V. P. Antropov, M. I. Katsnelson, M. van Schilfgaarde, and B. N. Harmon, Phys. Rev. B \textbf{75}, 729 (1995).
\bibitem{16.2Antropov} V. P. Antropov, M. I. Katsnelson, B. N. Harmon, M. van Schilfgaarde, and D. Kusnezov, Phys. Rev. B \textbf{54}, 1019 (1996).
\bibitem{16.3Savrasov} S. Y. Savrasov, Phys. Rev. Lett. \textbf{81}, 2570 (1998).
\bibitem{17Inoue} J. Inoue and S. Maekawa, Phys. Rev. Lett. \textbf{74}, 3407 (1995). 
\bibitem{18Perdew} J. P. Perdew, K. Burke, and M. Ernzerhof, Phys. Rev. Lett. \textbf{77}, 3865 (1996).  
\bibitem{19Ceperley} D. M. Ceperley and B. J. Alder, Phys. Rev. B \textbf{45}, 566 (1980).
\bibitem{Shallcross} S. Shallcross and A. E. Kissavos, Phys. Rev. B \textbf{72}, 104437 (2005). 
\bibitem{20Stringfellow} M. W. Stringfellow, J. Phys. C \textbf{1}, 950 (1968).
\bibitem{21Lynn} J. W. Lynn, Phys. Rev. B \textbf{11}, 2624 (1975).
\bibitem{21aPauthenet} R. Pauthenet, J. Appl. Phys. \textbf{53}, 8187 (1982). 
\bibitem{22Loong} C. K. Loong, J. M. Carpenter, J. W. Lynn, R. A. Robinson, and H. A. Mook, J. Appl. Phys. \textbf{55}, 1895 (1984).
\bibitem{Yoon} H. Yoon, T. J. Kim, J. H. Sim, S. W. Jang, T. Ozaki, and M. J. Han, Phys. Rev. B \textbf{97}, 125132 (2018).
\bibitem{nanorib1} Q. Yan, B. Huang, J. Yu, F. Zheng, J. Zang, J. Wu,
B. L. Gu, F. Liu, and W. Duan, Nano Lett. \textbf{7}, 1469 (2007).
\bibitem{nanorib2} D. Soriano, F. Mun$\tilde{\textrm{o}}$z-Rojas, J. Fern$\acute{\textrm{a}}$ndez-Rossier, and J. J. Palacios, Phys. Rev. B \textbf{81}, 165409 (2010).
\bibitem{30nanorib1} E. Rudberg, P. Sa\l{}ek, and Y. Luo, Nano Lett. \textbf{7}, 2211 (2007).
\bibitem{31nanorib2} E. Kan, Z. Li, J. Yang, and J. G. Hou, Appl. Phys. Lett. \textbf{91}, 243116 (2007). 
\bibitem{32nanorib3} S. B. Kumar and J. Guoa, Appl. Phys. Lett. \textbf{98}, 263105 (2011).
\bibitem{34nanorib5} K. Sawada, F. Ishii, M. Saito, S. Okada, and T. Kawai, Nano Lett. \textbf{9}, 269 (2009).
\bibitem{35nanorib6} K. Sawada, F. Ishii, and M. Saito, Appl. Phys. Lett. \textbf{104}, 143111 (2014). 
\bibitem{36nanorib7} D. M. Edwards and M. I. Katsnelson, J. Phys.: Condens. Matter \textbf{18}, 7209 (2006).
\bibitem{37nanorib8} O. V. Yazyev and M. I. Katsnelson, Phys. Rev. Lett. \textbf{100}, 047209 (2008).
\bibitem{38GBTnano} D. B. Zhang and S. H. Wei, npj Comput. Mater. \textbf{3}, 1 (2017).
\end{thebibliography}
\end{document}